# Optomechanical suspended waveguide for broadband phase modulation with frequency memory effect

Enrico Casamenti*, Tao Yang, Pieter Vlugter, and Yves Bellouard

*Galatea Laboratory, IMT/STI, Ecole Polytechnique Fédérale de Lausanne (EPFL), Rue de la Maladière 71b, CH - 2000 Neuchâtel,Switzerland*

*Corresponding author: enrico.casamenti@epfl.ch*

**Whether it is for transmitting information or for controlling intensity, light modulation is among the essential functions commonly used in complex optical systems. In integrated optics, modulation principles usually include the use of electro-optical effects or acousto-optics varying index waves. Here, we demonstrate a concept of light-modulation based on a vibrating suspended cantilever clamped on both ends, acting as an optical waveguide. In this approach, an optical phase shift is introduced by the physical elongation and by the stress induced in the vibrating cantilever. Most interesting, such a device can store optical information in the form of a phase shift, related to the frequency path-dependence of the vibration. This remarkable property stems from the intrinsic non-linear dynamical behavior of the device, resembling Duffing-like oscillators. In this proof-of-concept, the micro-device is manufactured in a single fused silica chip using femtosecond laser exposure combined with chemical etching and $CO_2$-laser polishing to achieve optical roughness quality. The cantilever vibration results from dielectrophoresis forces arising in the silica beam as a consequence of the non-uniform electrostatic field applied to it. Noticeably, being made entirely of fused silica, the device is capable of broadband operation, from the near UV to the mid-IR range, and this, at kHz frequencies.**

## 1. Introduction

Modulating light in a controlled manner, whether it is in phase or intensity is among the basic function commonly found in optical systems and serves many different purposes. From transmitting data to controlling laser exposure conditions or from optical metrology to signal conditioning in general, light modulation is ubiquitous. In this work, we focus our attention to phase modulation principles due to their broad use and applicability for integrated optics applications.

A phase modulator is a device capable of dynamically tune the phase shift induced to a light beam traveling through it. Depending on the application, there are few key metrics to consider when dealing with phase modulation. One is the achievable maximum phase shift, usually known as the modulation index. Others are the driving voltage required by the device, the modulation frequency capability, the optical bandwidth and for integrated optics in particular, the dimensions of the device and its scalability.

In most devices, phase modulation is usually achieved by tuning the dielectric response (and therefore the refractive index) of the material. Using electro-optics principles such as the Pockels effect, modulations up to GHz frequencies have been demonstrated in polymers [1], graphene [2] and lithium niobate [3]. Other approaches manipulate the refractive index of a medium by temperature changes [4] or by indirect means based on piezoelectric transducer to create traveling stress waves in the medium and in turn, to achieve an acousto-optic modulator [5].

Following a global trend towards systems miniaturization and integration in photonics, miniature optical modulators have recently been introduced, like for instance systems based on lithium niobate crystals ($LiNbO_3$) [3]. These devices exploit the large Pockels effect observed in these crystals (i.e. $r_{33}$ ~ 30 pm/V) [5], but are limited by the plating of electrodes to two-dimensional space, which restricts their integration capability. Despite a broad optical spectrum (ranging approximately from 600 to 3500 nm), Lithium-Niobate transmission performance remains below 80%. Optical substrate like silica has excellent optical properties, however, being centrosymmetric, has no second-order nonlinearity. This limitation can be partially overcome using an electrothermal post-processing known as 'thermal polling'. Such process breaks the centrosymmetry of the glass [6] and induces a non-zero value of Pockels coefficient. Commonly investigated for fibers, this approach has also been proposed by Winick *et al.* to build an optical modulator fully integrated into a fused silica substrate containing waveguides. [7] However, the device presented in their work, due to the low induced nonlinearity, possesses a modulation index of only 1 nm approximately, for a driving voltage of 400 V and an interaction length of 25.6 mm (more than two orders of magnitude less than an equivalent $LiNbO_3$ substrate).

The objective of our work is to explore a new approach to achieve optical modulation in a single fused silica substrate. Our approach, based on optomechanics, is to use a suspended waveguide put under vibration. The device, due to oscillations, imposes a controllable phase shift by a combination of physical elongation and photoelasticity, i.e. stress-induced retardance. Such an integrated phase modulator, thanks to its intrinsic nonlinear dynamics, can also operate as a switch, a filter, and a passive vibration fault all-optical sensor.

## 2. Working principle, fabrication and modeling

### A. Working principle

The device concept is an optically transparent suspended beam, clamped at both ends, that is set in resonance to induce a phase modulation of an arbitrary optical wave passing through. In the case of electrostatic actuation, the device effectively transduces an electrical signal into an optical one. Noticeably, as a direct consequence of the double-clamped boundary condition, the vibration dynamical response is nonlinear. In particular, it exhibits a signature hysteresis in the frequency domain. Interestingly, this frequency path dependence to the oscillations amplitude creates a 'memory effect': the device can intrinsically 'remember' whether it was subjected to a frequency increase or decrease. This remarkable property opens up interesting perspectives for applications that go beyond optical modulation.

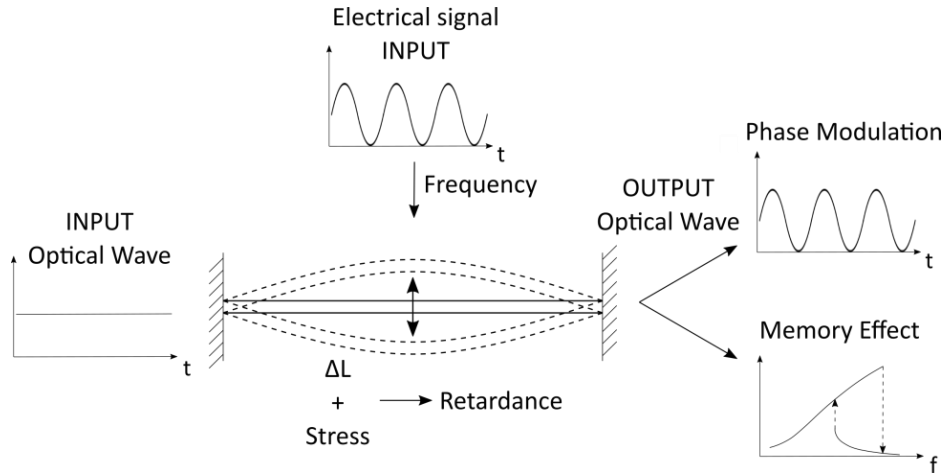

**Fig. 1.** Device concept: working principle. A transparent vibrating suspended beam causes an optical modulation of the optical wave traveling through, whose modulation amplitude is not only proportional to the vibration amplitude, but also to the frequency path history (*memory effect*).

**B. Proof-of-concept**

To demonstrate the concept discussed in the previous section, we designed a suspended double-clamped fused silica cantilever placed in a V-shaped cavity [8]. There, the beam is set into vibration using dielectrophoresis – a second-order electrostatic effect that can be used to induce a force-field on a dielectric object. Through oscillations, the device modulates the phase of an optical wave traveling in it due to the effective change of length and of the stress induced in the beam. A schematic of the proof-of-concept device is shown below and an optical view-graph in the figure thereafter.

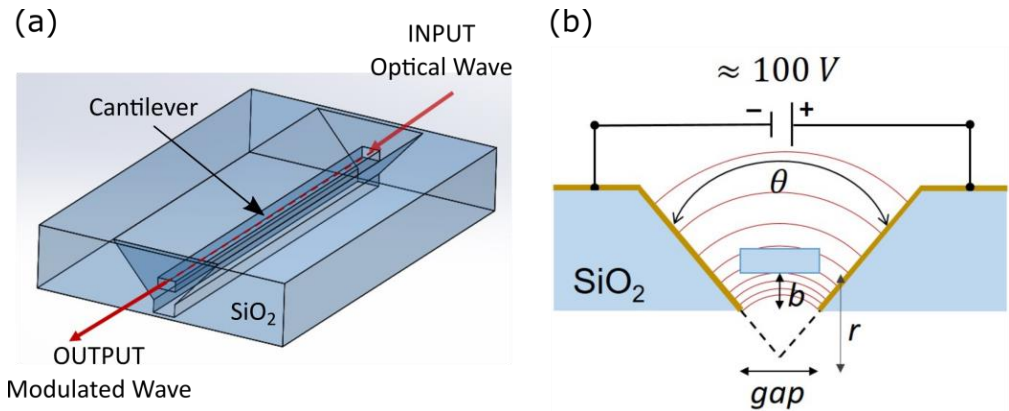

**Fig. 2.** (a) Proof-of-concept silica-based optical modulator. (b) Cross-section and key design parameters used to optimize dielectrophoresis-based actuation. The yellow line illustrates where gold is deposited to form electrodes.

With this technological approach, we predict to achieve a phase modulator with a large modulation index, a low driving voltage (< 100 V) compared to electro-optical modulators and fully integrated into fused silica to achieve broadband operation (approximately from 0.2 to 2 µm).

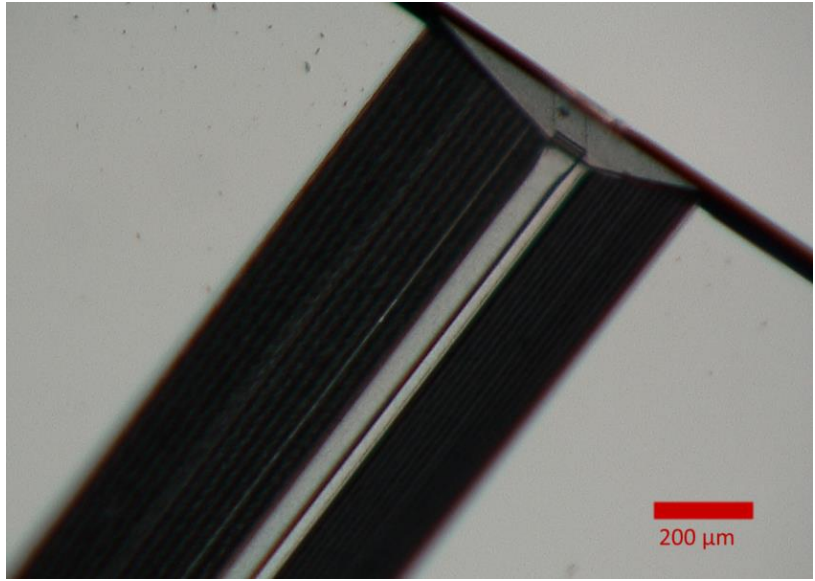

**Fig. 3.** Optical microscope picture showing one side of the prototype before plating the V-groove with gold. The cantilever was exposed to a CO$_2$ laser to achieve optical quality surfaces.

## B. Fabrication

The device is fabricated using femtosecond laser micromachining followed by wet chemical etching, according to a process described elsewhere [9]. In the present study, we used a Ytterbium-fiber amplifier laser emitting 270 fs pulses at a repetition rate of 800 kHz, with an energy of 240 nJ. The light beam is focused by a 0.4-numerical aperture objective to a waist of 1.5 µm. The substrate used is a UV fused silica (Corning 7980 0F), 500 µm-thick, 25 mm-squared. The specimen is loaded on a two-axis motorized stage, enabling planar motion, whereas the laser focal spot is translated along the vertical direction through a linear stage. Following laser exposure, the specimen is etched in a 2.5 % HF acid bath for several hours.

In the present study, the cantilever is 23 mm long and has a 50 x 100 µm² cross-section. A final step of CO$_2$-laser polishing is implemented to achieve optical surface quality and in turn, to limit scattering losses. [10] Finally, a thin gold layer is sputter-deposited on the V-groove and through a glass mask to form two electrodes, while preventing deposition of gold on the cantilever.

## C. Actuation modeling

In earlier works, T. Yang *et al.* [8,11] modeled free-standing and double-clamped beams under dielectrophoresis (DEP) actuation. In the sequel, we briefly summarize the main results. Based on this model, the DEP force writes:

$$\boldsymbol{F}_{DEP} = C_v \varepsilon_m \operatorname{Re}(K) \nabla \boldsymbol{E}^2 \qquad (1)$$

Where $C_v$ is the volume of the object, $\varepsilon_m$ is the permittivity of the medium, $\operatorname{Re}(K)$ is the real part of the Clausius-Mossotti factor $K$ and $E$ is the applied electric field. For a non-spherical object, a form factor is implemented to correct the C-M factor. In the simple case of a V-shaped electrodes configuration, the expression of the electrical field gradient writes :

$$|\nabla E^2| = 2V^2 r^{-3} \theta^{-2} \qquad (2)$$

where $V$ is the applied voltage, $r$ is the distance between the cantilever and the virtual tip of the V-groove and $\theta$ is the opening angle of the V-groove (see Fig. 2b). Assuming the two electrodes as infinite long planes along the beam direction, the force is proportional to three key design parameters as follows [8]:

$$|F_{DEP}| \propto \left(b + \frac{gap}{2\tan(\theta/2)}\right)^{-3} \theta^{-2} \qquad (3)$$

After performing an optimization study, the following values were selected: $b = 50$ µm $gap = 120$ µm and $\theta = 90°$.

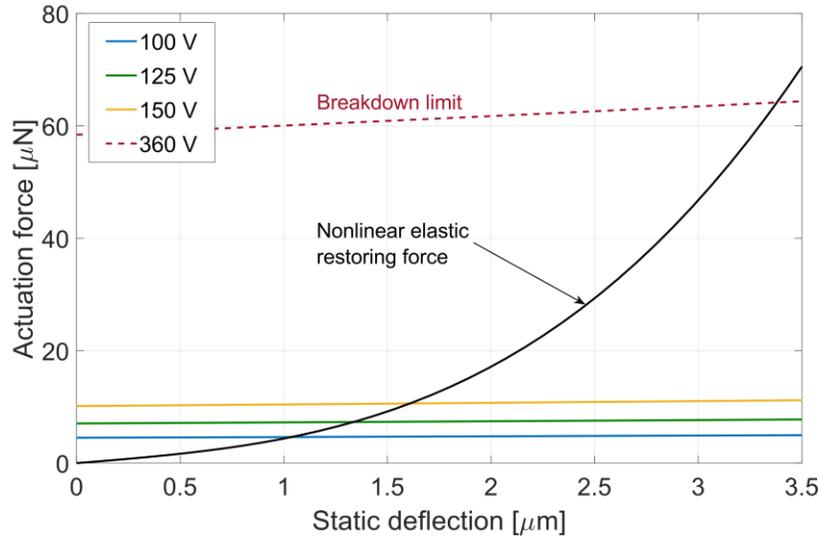

**Fig. 4.** A plot of the dielectrophoresis actuation force versus the deflection of the beam in the static case for the design values mentioned above. The intersections with the nonlinear elastic restoring force represent the equilibrium points. The breakdown voltage limit is also shown.

A clamped-clamped condition for an oscillating beam has an intrinsic non-linear dynamical behavior. Indeed, as the beam deflects, a tension proportional to the beam deflection arises. An analytical description to capture this dynamic behavior is performed adopting a Duffing-like oscillator model [12] as described in [11]. This leads to the harmonic equation:

$$\frac{d^2 y(t)}{dt^2} + 2\delta \frac{dy(t)}{dt} + \alpha_1 y(t) + \alpha_2 y^2(t) + \alpha_3 y^3(t) = b_1 + b_2 \cos(\omega t) \qquad (4)$$

where $y$ is the vertical displacement, $2\delta \, dy(t)/dt$ is a dissipation term, $b_1$ is the static excitation, $b_2 \cos(\omega t)$ is the harmonic excitation and $\alpha_1$, $\alpha_2$ and $\alpha_3$ are the linear, quadratic and cubic mass normalized nonlinear stiffness coefficients, respectively. Note that in practice the excitation term is made of two terms. One DC component and one AC term. In the static case, comparing the DEP actuation magnitude with the non-linear elastic restoring force, we show a deflection of around 1 µm with a driving voltage close to 100 V. In addition, from Euler-Bernoulli beam theory and finite element analysis, the frequency response is estimated, with the first natural frequency found around $f_1 = 612$ Hz for this particular case.

## 3. Experimental set-up

### A. Actuation

To actuate the device a DC+AC voltage is applied (Rigol DG1032 signal generator combined with a Tabor Electronics 9200A amplifier). To measure the beam displacement a laser triangulation sensor (Keyence LK-H022K) is used. Different combinations of voltages are tested, with the DC voltage ranging from 50 to 150 V for the static loading and an AC ranging from 25 to 200 V peak-to-peak amplitude. The dynamic behavior is investigated by performing various frequency sweeps cycles around the first resonance and using a lock-in amplifier (Stanford Research Systems SR830 DSP).

### B. Phase measurement

A 632 nm-laser is injected in the suspended waveguide using standard free space optics and collected using a microscope objective. Phase shifting in the optical signal due to the beam deflection is measured using a Mach-Zehnder Interferometer (MZI) (see Fig. 5). A delay line is arranged to account for the difference in path lengths between the two arms. Finally, standard CMOS camera (Thorlabs DCC1645C) and photodetector (Thorlabs PDA36A2) are used to finalize the alignment and to measure the intensity of the light beam after interference, respectively.

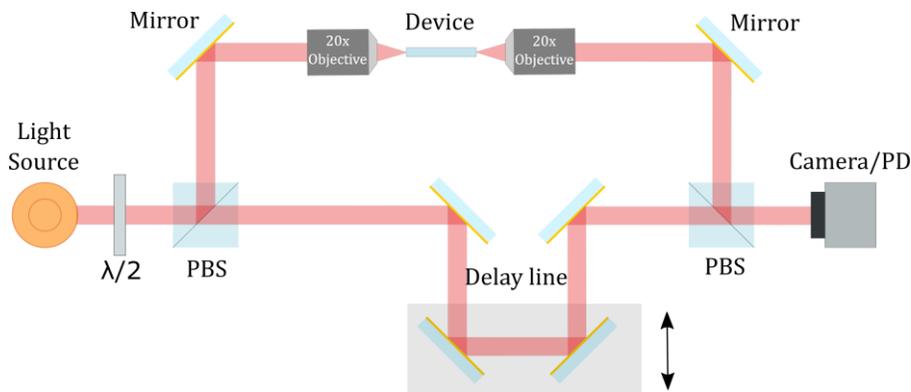

**Fig. 5.** Schematic of the optical phase shit measurement setup (Mach-Zehnder interferometer combined with a delay line).

## 4. Results

Figure 6 shows experimental results of the beam vibration amplitude as a function of the frequency and the actuating voltage. The first natural frequency peak is found at 609 Hz for an actuation voltage of 100 V in DC plus 25 V in AC peak-to-peak and it corresponds to a maximum deflection of 18 µm, which corresponds to around 0.08 % of the length of the beam and about one-third of the beam thickness. Further, the same hardening nonlinearity already described in [11] is monitored with the displacement detector. In particular, for increasing value of AC voltage over the DC, the difference in behavior between sweeping up or down in frequency becomes pronounced, exhibiting a strong hysteresis. A maximum displacement of more than 80 µm is reported at mid-length of the cantilever when the device oscillates at its first resonance frequency with an actuation voltage of 100 V in DC plus 200 V in AC peak-to-peak. Increasing the actuation voltage, either AC or DC, causes a positive shift of the resonance peaks. Differently, no changes are recorded for the Q-factor, which remains around 50 in standard laboratory conditions. We note also a small path dependence at low oscillation amplitudes which is not to be expected according to the Duffing oscillator model used here and further detailed in [11]. This effect might be related to further non-linearities in the system, for instance, caused by imperfect cross-section shape of the beam or by the excitation of super- or sub- harmonics.

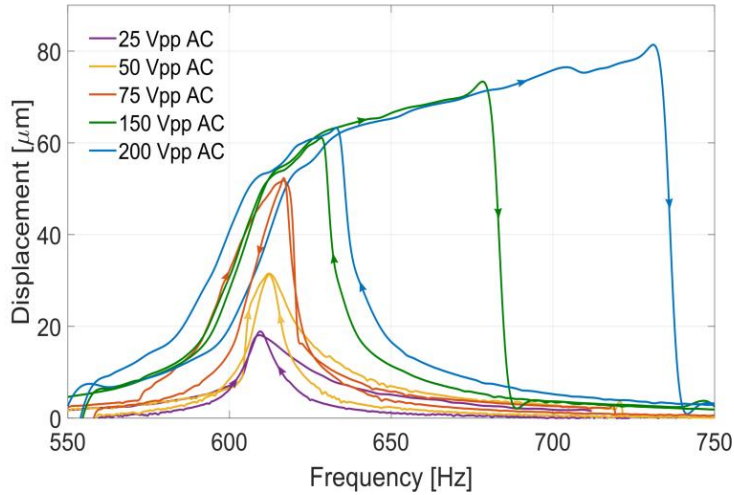

**Fig. 6.** Nonlinear dynamic response of the double-clamped cantilever actuated by dielectrophoresis. The device is under a 100 V static voltage onto which an oscillating voltage of varying intensity is superposed in order to set the beam in vibration. Dynamic behavior for both sweeping up and down in frequency are shown, exhibiting a strong hysteresis and hardening behavior. Displacement error measurement is +/- 0.1 micron, while the frequency error measurement is estimated to be +/- 2 Hz.

From the displacement measured in the center of the cantilever, using classical Euler-Bernoulli beam theory, we retrieve an equivalent uniformly distributed load:

$$P = \frac{y}{\frac{1}{24EJ_x}\left(-z^4 + 2Lz^3 - L^2z^2\right)} \qquad (5)$$

where $y$ is the vertical displacement, E is the Young modulus of fused silica, $J_x = t^3w/12$ is the second moment of inertia (with $t$ the thickness and $w$ the width of the cantilever cross-section), $z$ the position along the cantilever's length, and finally, $L$ the length of the cantilever. Note that as we are considering the first vibration mode, the maximum displacement is assumed to be exactly at $z=L/2$. The total cantilever elongation resulting from its deformation is obtained by integrating the beam profile along its length. For a given distributed load, the beam profile for a downwards deformation is written as:

$$y(z) = \frac{P}{24EJ_x}\left(-z^4 + 2Lz^3 - L^2z^2\right) \quad (6)$$

Thus, by inserting Eq. 5 in Eq. 6, the elongation $\Delta L$ writes:

$$\Delta L = \int_0^L \sqrt{1 + (dy/dz)^2}\, dz - L \quad (7)$$

As an illustration, according to this model, for a maximum deflection of 33 μm the 23 mm-long cantilever elongates by 115 nm. For the largest maximum deflection recorded (82 μm) in this experiment, the elongation reaches 712 nm. When considering the optical signal traveling in the cantilever, this physical elongation is directly related to a change in optical path length. Therefore, beam elongation is the primary source of phase shifting induced by the device when actuated.

Furthermore, the deflection of the cantilever induces stress in the beam, which in turn modulates the refractive index, due to photoelasticity, and consequently, the optical phase retardance. The clamped-clamped boundary conditions cause both bending and tensile stresses to act along the beam length. As one can see using finite element models, the stress is prevalently in the direction of beam length where the combined effect of tension and bending accounts for a few MPa, which remains very low compared to the maximum stress that the glass can withstand, which for fused silica, can exceed a few GPa. As an illustration, for 33 μm of maximum deflection, the maximum stress observed near the anchoring points is 4.5 MPa, while the average stress accounts for 1 MPa. However, since the cantilever is 23 mm long, the stress-induced phase retardance is not negligible, albeit smaller than the one induced by the change of path-length. To estimate it, the change in refractive index $\Delta n$ is integrated along the entire length of the cantilever:

$$\delta = \int_0^L \Delta n \sqrt{1 + (dy/dz)^2}\, dz \quad (8)$$

with $\delta$ the induced retardance. Assuming $\Delta n = CT^z$, where $C = 3.55 \cdot 10^{-12}\,\text{Pa}^{-1}$ is the relevant stress-optic coefficient for fused silica and $T^z$ is the average stress along the direction of propagation of light, estimated from a simple FEM model. It should be noticed that in this case, the stress state perpendicular to the propagation of light that induces birefringence is negligible. In fact, the $T^x$ and $T^y$ values are mostly similar along the length of the cantilever and are in the order of $10^{-1}$ MPa. In what follows, we only account for the phase retardance induced along the propagation of light. From the analytical model, a maximum deflection of 33 μm results in a stress-induced retardance of around 29 nm. Thus, summing the phase shift due to the elongation of the beam with the one induced by the stress, for 33 μm of maximum deflection we predict a total retardance $\delta_{tot} \approx 144$ nm.

To investigate the optical modulation, a monochromatic light at 632 nm is launched in the cantilever and collimated at the sample output. Then it interferes with a second beam used as a reference as shown in Fig. 5. The device is actuated applying 100 V DC plus different values of AC voltage to stimulate the oscillation. The normalized intensity signals from the photodetector are shown in Fig. 7. The increase in modulation related to the increasing AC actuating voltage is observed.

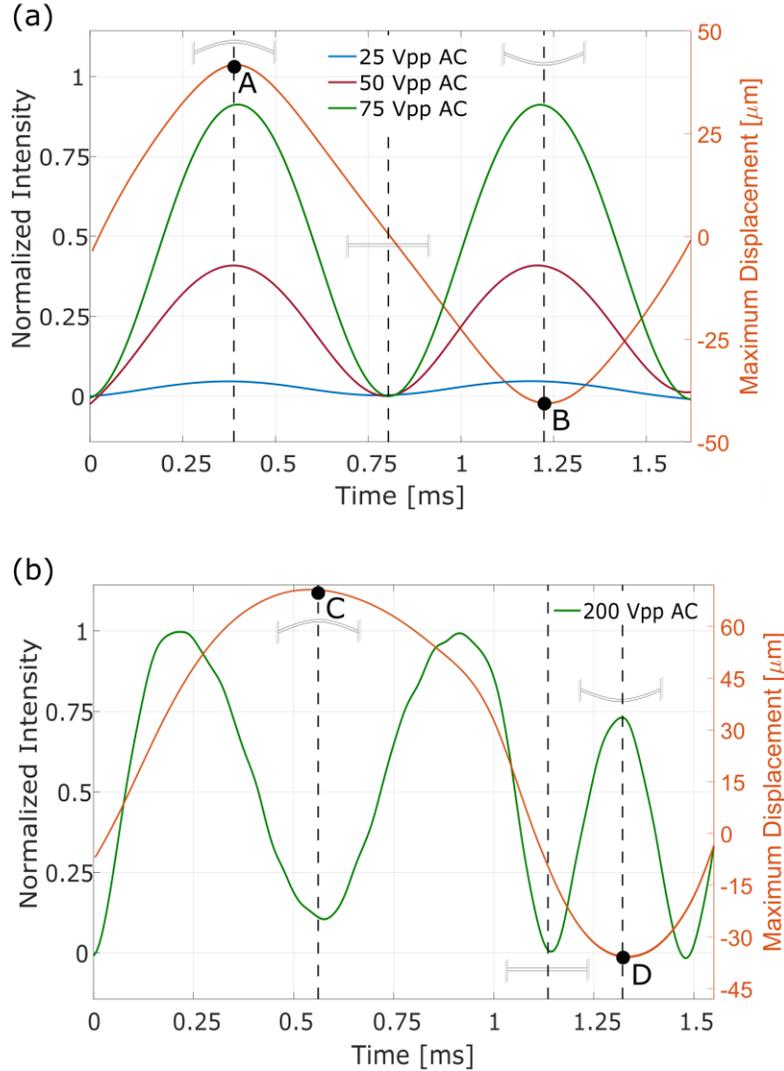

**Fig. 7.** Photodetector intensity modulated signal with an actuation voltage of 100 V DC carrying a varying AC voltage at different frequencies. The signal is normalized with respect to the maximum of intensity observed in Fig. 7b. The displacement of the double-clamped cantilever is also plotted. (a) Case where the induced retardance is lower than the modulated light half-wavelength. The three measurements are performed at a fixed frequency of 620 Hz. (b) Case for which the induced retardance becomes larger than the half wavelength. After a positive sweeping from 600 to 680 Hz, the frequency is kept fixed at 680 Hz. A phase wrapping is clearly visible along with an asymmetric behavior.

In Fig. 7a, the device is actuated with an AC signal at a constant frequency of 620 Hz. In Fig. 7b, the frequency is incremented performing a positive sweep from 600 Hz until 680 Hz to increase the cantilever deflection, entering the non-linear regime in a known state (see Fig. 6). For the measurement, the frequency is then kept constant at 680 Hz.

Let us first examine the case of Fig. 7a. Due to the intrinsic working principle of the device, we observe two maximum deflection positions, one up (A) and one down (B), further shown in Fig. 8. As a consequence, we obtain a modulation at twice the frequency of the actuating signal (see Fig. 7a). Further increasing the voltage and the frequency reveals optical phase-wrapping, as can be seen in Fig. b. Likewise, we obtain a double phase-wrapping, both on the high and low peak.

However, we notice an asymmetry in the oscillating position of the cantilever, when the AC voltage is increased (Fig. 7b). We interpret this asymmetry as follows (see Fig. 8): as the cantilever oscillation amplitude increases, it eventually reaches a downward position approaching the actual edges of the V-grooves (position D, in Fig. 8 iii). There, the gap between the cantilever and the V-groove closes to a few microns, dramatically increasing the air resistance [13] and the DEP forces (that scales with the inverse of the cube of separation distance). However, as the cantilever approaches the V-groove edges, the fringes effects in the electrostatic field due to the sharp V-groove corners become dominant. This modifies the DEP force orientation – which is parallel to the field gradient – and it eventually works against the downward motion, preventing the beam from moving further down (Fig. 8 iii). This asymmetry between upward and downward movements is further amplified by the damping effect (proportional to the speed) of the squeezed thin air film at the bottom of the V-groove as the cantilever closes the gap while moving down.

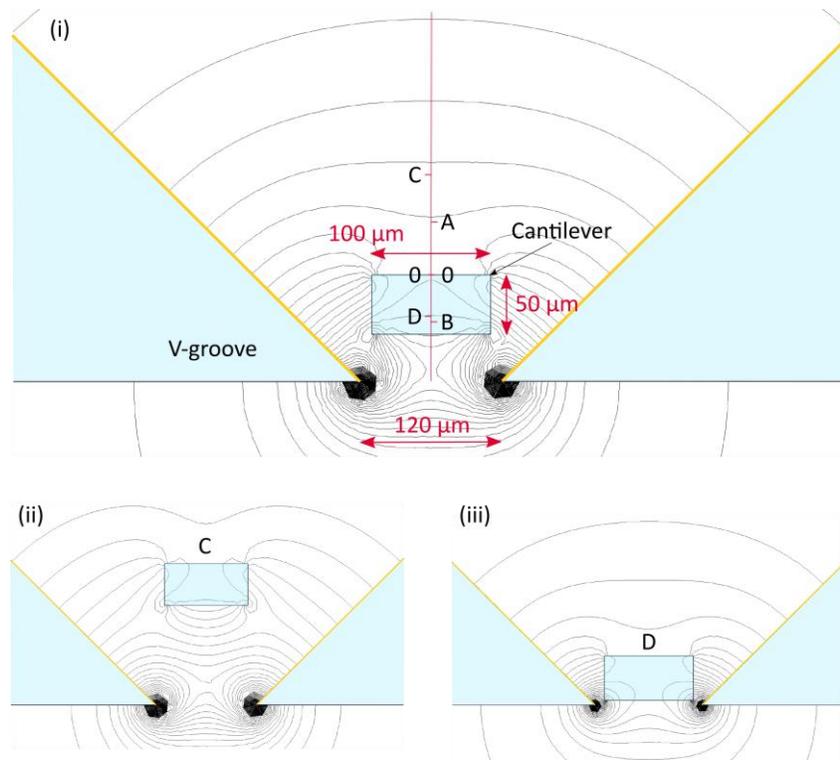

**Fig. 8.** Drawing of the cross section of the device in its middle part – i.e. where it deflects the most, and where the measurements were made. The actual position A, B, C & D correspond to the specific points in the curves shown in Fig. 7. The simulated electrical field distribution is shown with equipotential lines (in black). In this FEM simulation, a potential of 200 V is imposed between the two electrodes (represented by the gold lines). The structure is assumed to be infinite in the out-of-plane direction, so that the problem can be reduced to a two-dimensional one. As an approximation, we assume a perfectly rectangular cross-section. (ii) and (iii) illustrate the particular case of position C and D.

To explore the memory effect on the optical modulation, the same set of actuating voltages used in Fig.7a, is applied to the device, but this time with a frequency sweep, ranging from 550 to 750 Hz, back and forth. The results are shown in Fig. . Similarities with the nonlinear behavior of the beam displacement amplitude shown in Fig. 6 are observed and manifest themselves by a difference in both, amplitude and frequency, visible while sweeping toward higher or lower frequencies. In this case, the dynamic response for actuating AC voltages above 90 V is not shown. Above this limit, the wrapping of the intensity signal starts to happen as illustrated in Fig. 7, disturbing the measurements of the lock-in amplifier with a higher frequency signal component. Nevertheless, the nonlinearity of the dynamic response can already be observed with an AC actuating voltage of 50 and 75 $V_{pp}$, which further indicates that the observed non-linearity at low amplitude of vibrations, albeit small and that may be mistaken for measurement artefacts, are effectively present. Interestingly, these non-linearity at low amplitude are not captured by the theoretical model based on a pure Duffing oscillator, presented in a previous study [11].

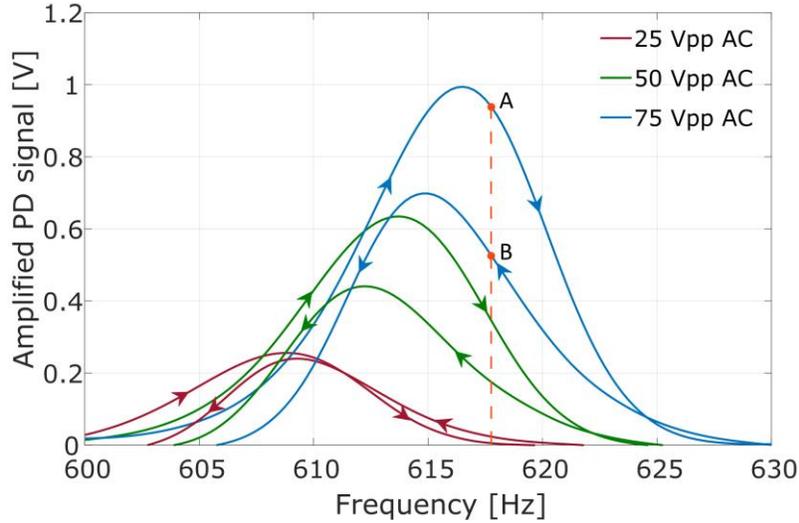

**Fig. 9.** Photodetector amplified intensity signal as a function of frequency (sweeping up and down). The device is driven at 100 V DC onto which is added an AC voltages of various amplitudes. The sweepings range from 550 to 750 Hz, with an overall sweeping time of 10 seconds.

This nonlinear behavior indicates a 'memory effect' of the device: the amplitude of the modulation *depends* on the direction of the frequency sweep (see for instance Fig. , points A and B). As a direct consequence, by processing the photodetector output signal, it is possible to retrieve the previous frequency state of the cantilever beam. In other words, the device can 'remember' whether it came from higher or lower frequencies.

The final step is to compare the experimental data of the modulated intensity signal with the theoretical model to retrieve the value of the induced phase shift. Assuming the intensity of the light beams of the two arms of the Mach-Zehnder interferometer are balanced (i.e. $I_1 = I_2 = I_0$), we can express the intensity after interference as $I_{tot} = 2I_0 \{1 + \cos[\varphi(t)]\}$. In particular, the phase difference can be described as $\varphi(t) = \Delta\varphi_{max} \sin(\omega t)$, where $\Delta\varphi_{max}$ is the maximum phase shift induced by the device (i.e. the one corresponding to the maximum deflection of the cantilever) and $\sin(\omega t)$ is the sinusoidal oscillation of the cantilever at its first resonance frequency. For instance, by comparing the model to the experimental data, for an actuation of 100 V DC plus 50 $V_{pp}$ AC, the maximum phase shift is calculated to be $\Delta\varphi_{max} = 0.46\pi$, which corresponds to a phase retardance of 145 nm at a wavelength of 632 nm, in good agreement with the value predicted by the analytical model.

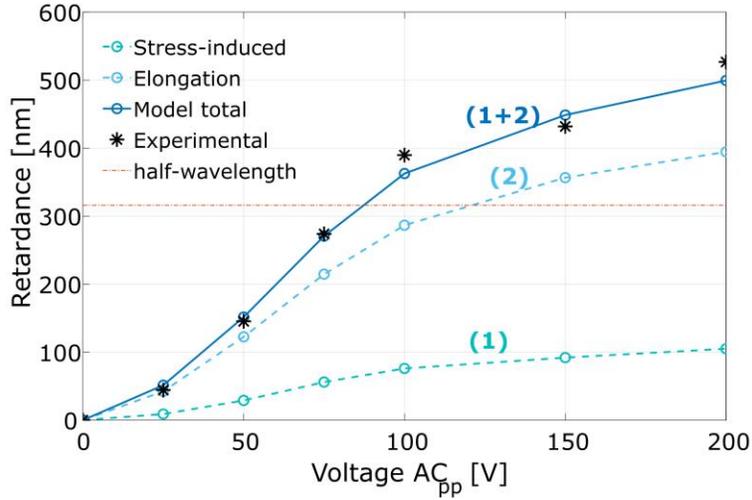

**Fig. 10.** Induced optical phase retardance and maximum upward deflection versus AC actuating voltage at the resonant frequency (a constant 100 V DC is added). The experimental results (in black – for the maximum optical phase retardance) and the model (in blue – for the deflection), with the contribution from stress and elongation (in turquoise), are represented. Connected lines are shown as a visual aid. Experimental error bars are estimated to be +/- 10 nm for the retardance.

By increasing the AC driving voltage up to 200 $V_{pp}$, it is possible to reach values of total induced retardance above 500 nm. In the design considered here (with electrodes 120 μm apart at minimum), the maximum voltage limit is defined by the air breakdown that limits the actuation voltage to 360 V at maximum. Such an actuating voltage will result in a displacement of about 160 μm, approximately twice larger than the maximum used in our intensity modulation experiment, which would induce a path length difference of about 2713 nm (Eq. 7), exceeding the spectral bandwidth of fused silica.

## 5. Discussion and scaling analysis

As demonstrated here, in this modulation principle, the optical phase shift is mainly due to the physical elongation of the cantilever and to a lesser extent to stress-induced retardance. This simple working principle is generic and can be applied to any optical wavelength within the optical transparency spectrum of the raw material. For fused silica, this optimal optical transparency window typically spans from about 180 nm (depending on the material grade) to roughly 2 microns. This opens to full modulation from the UV up to the near-infrared spectrum.

The modulation frequencies – in this design example around 600 until 700 Hz – can be further increased to the kHz region, by either operating the device at higher harmonics of the natural frequency or by reducing its size.

In this respect, Fig. below illustrates how the design can be scaled down to achieve high operating frequencies for a given spectral modulation band. There, the ratio between height, width, and length of the cantilever is maintained constant as illustrated in the figure. In this simulation, the actuating voltage is scaled down proportionally to the dimensions. By assuming a device operation under low vacuum (about 1 mBar) and considering the negligible internal friction of fused silica, the Q-factor is estimated to be 1000. [13] Using the model presented in the previous section, the static displacement is found and, considering the dynamic case, the elongation is computed with Eq. 7. In this analysis, we neglect stress-induced retardance. Doing so, we represent a less favorable scenario as it overestimates the required actuating voltage.

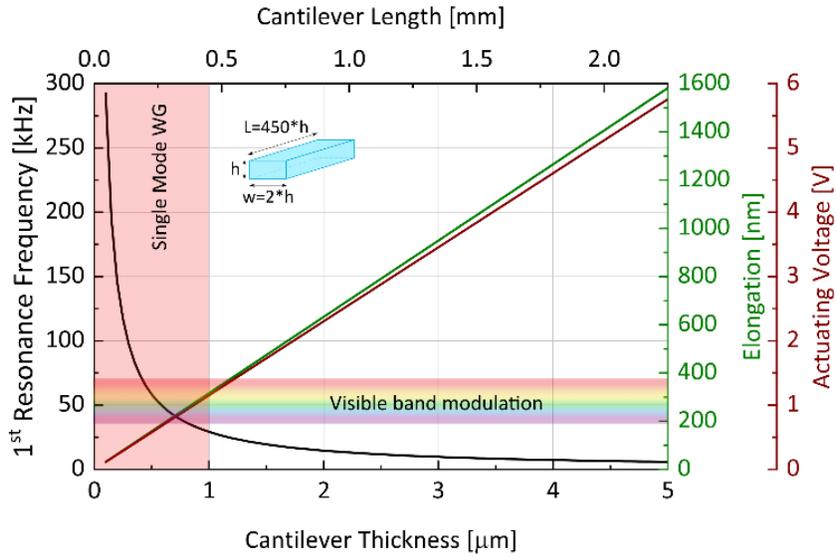

**Fig. 11.** Scaling effect on the resonant frequency, voltage, and optical modulation band. The main design parameter here is the cantilever thickness, all the other dimensions being directly coupled through a proportionality relation. The visible band spectrum shows the requirements to obtain a full modulation. The region in red on the left is added to visualize the dimensions limit to obtain single mode waveguiding in the beam, within the transmission spectrum of fused silica.

This simple analysis shows that scaling down the device does not reduce the modulation bandwidth, while it largely increases the frequency of operation and decreases significantly the actuating voltage required. It demonstrates the potential of such a device for broadband optical modulation at low-voltage and tens of kHz frequency bandwidth.

## 6. Conclusion and outlook

To summarize, we presented a proof-of-concept monolithic optomechanical phase modulator integrated into fused silica and manufactured out of a single substrate using femtosecond laser exposure combined with chemical etching. The working principle is to couple the vibration of a double-clamped transparent beam with an optical beam and to achieve phase modulation by dynamically varying the optical path length. Intrinsically simple, thanks to its wavelength independency, the device can be operated within a broad optical spectrum, corresponding to the optical transparency window of the substrate. Here, a maximum induced optical phase retardance of about 530 nm has been demonstrated experimentally using DEP actuation (for an actuating voltage of 100 V in DC with a modulated voltage of 200 $V_{pp}$ in AC) and at a frequency of about 700 Hz. A simple scaling analysis demonstrates that the concept could operate at a much lower voltage (a few V) and up to tens of kHz of operating frequency, with a maximum phase shift covering the entire fused silica's transmittance spectrum.

Furthermore, such a device exhibits an intrinsic nonlinear dynamical response with a characteristic path-dependency on the frequency sweeping direction. This effect can be seen as the device recording its previous frequency states. This offers an interesting perspective for applications beyond optical phase modulation, such as non-linear signal processing. Note that in another work, we demonstrated that this characteristic frequency response can be further tuned by re-exposing the beam to femtosecond laser irradiation [11].

Finally, the device offers a mean for converting a mechanical vibration directly into an optical one, without the need for electrical transducing. This is particularly attractive for harsh environment all-optical sensing, such as vibration fault detection in aeronautics.

**Funding**. The Galatea Lab acknowledges the sponsoring of Richemont International.

**Acknowledgment.** In this work, E.C., wrote the draft manuscript, implemented the optical modulation concept, modeled it and carried out the experiments, T. Y. designed the proof-of-concept and modelled the DEP actuation and mechanical response, P. V. contributed to the vibration dynamic characterization of the device, Y. B. proposed the initial concept and supervised the research. The authors further thanks Dr. S. Hakobyan for the fruitful discussions and Dr. J. Gateau for the help in the implementation of the MZI set-up.

# REFERENCES

1. J. Liu, G. Xu, F. Liu, I. Kityk, and Z. Zhen, "Recent advances in polymer electro-optic," RSC Adv. **5**, 15784–15794 (2015).
2. C. T. Phare and J. C. and M. L. , Yoon-Ho Daniel Lee, "Graphene electro-optic modulator with 30 GHz bandwidth," Nat. Photonics **9**, 511–515 (2015).
3. D. Janner, D. Tulli, and M. Garc, "Micro-structured integrated electro-optic LiNbO 3 modulators," **313**, 301–313 (2009).
4. J. W. Silverstone, D. Bonneau, K. Ohira, N. Suzuki, H. Yoshida, N. Iizuka, M. Ezaki, C. M. Natarajan, M. G. Tanner, R. H. Hadfield, V. Zwiller, G. D. Marshall, J. G. Rarity, J. L. O. Brien, and M. G. Thompson, "On-chip quantum interference between silicon photon-pair sources," Nat. Photonics **8**, 104–108 (2013).
5. B. E. A. Saleh and M. C. Teich, *Fundamentals of Photonics*, Second Ed. (Wiley, 2006).
6. R. A. Myers, N. Mukherjee, and S. R. J. Brueck, "Large second-order nonlinearity in poled fused silica," **16**, 1732–1734 (1991).
7. G. Li, K. A. Winick, A. A. Said, M. Dugan, and P. Bado, "Waveguide electro-optic modulator in fused silica fabricated by femtosecond laser direct writing and thermal poling," Opt. Lett. **31**, 739–741 (2006).
8. T. Yang and Y. Bellouard, "Monolithic transparent 3D dielectrophoretic micro-actuator fabricated by femtosecond laser," J. Micromechanics Microengineering **25**, 105009 (8pp) (2015).
9. Y. Bellouard, A. Said, M. Dugan, and P. Bado, "Fabrication of high-aspect ratio, micro-fluidic channels and tunnels using femtosecond laser pulses and chemical etching," Opt. Express **12**, 2120–2129 (2004).
10. J. Drs, T. Kishi, and Y. Bellouard, "Laser-assisted morphing of complex three dimensional objects," Opt. Express **23**, 17355–17366 (2015).
11. T. Yang and Y. Bellouard, "Laser-Induced Transition between Nonlinear and Linear Resonant Behaviors of a Micromechanical Oscillator," Phys. Rev. Appl. **7**, 064002 (6pp) (2017).
12. I. Kovacic and M. J. Brennan, *The Duffing Equation Nonlinear Oscillators and Their Behaviour* (Wiley, 2011).
13. S. Schmid, L. G. Villanueva, and M. L. Roukes, *Fundamentals of Nanomechanical Resonators* (Springer, 2006).